\input harvmac
\def\journal#1&#2(#3){\unskip, \sl #1\ \bf #2 \rm(19#3) }
\def\andjournal#1&#2(#3){\sl #1~\bf #2 \rm (19#3) }

\def\ie{{\it i.e.}}
\def\eg{{\it e.g.}}

\def\frac#1#2{{#1\over#2}}

\def\inbar{\,\vrule height1.5ex width.4pt depth0pt}
\def\IC{\relax\hbox{$\inbar\kern-.3em{\rm C}$}}
\def\IR{\relax{\rm I\kern-.18em R}}
\def\IP{\relax{\rm I\kern-.18em P}}
\def\IZ{\relax{\rm I\kern-.18em Z}}

%
%

%
\catcode`\@=11
\def\slash#1{\mathord{\mathpalette\c@ncel{#1}}}\overfullrule=0pt

\def\SS{{\cal S}}

\def\underrel#1\over#2{\mathrel{\mathop{\kern\z@#1}\limits_{#2}}}

\catcode`\@=12

%

\def\det{{\rm det}}

\def \sinh{{\rm sinh}}\def \cosh{{\rm cosh}}

\def\det{{\rm det}}
\def\exp{{\rm exp}}


\lref\KutasovER{
D.~Kutasov and V.~Niarchos,
``Tachyon effective actions in open string theory,''
Nucl.\ Phys.\ B {\bf 666}, 56 (2003)
[arXiv:hep-th/0304045].
}

\lref\NiarchosRW{
V.~Niarchos,
``Notes on tachyon effective actions and Veneziano amplitudes,''
hep-th/0401066.
}

\lref\LukyanovNJ{
S.~L.~Lukyanov, E.~S.~Vitchev and A.~B.~Zamolodchikov,
``Integrable model of boundary interaction: The paperclip,''
arXiv:hep-th/0312168.
}

\lref\CallanAT{
C.~G.~.~Callan, J.~A.~Harvey and A.~Strominger,
``Supersymmetric string solitons,''
arXiv:hep-th/9112030.
}

\lref\SenAN{
A.~Sen,
``Field theory of tachyon matter,''
Mod.\ Phys.\ Lett.\ A {\bf 17}, 1797 (2002)
[arXiv:hep-th/0204143].
}

\lref\PolchinskiRQ{
J.~Polchinski,
``String Theory. Vol. 1: An Introduction To The Bosonic String,''
Cambridge University Press, 1998.
}

\lref\PolchinskiRR{
J.~Polchinski,
``String Theory. Vol. 2: Superstring Theory And Beyond,''
Cambridge University Press, 1998.
}

\lref\TseytlinDJ{
A.~A.~Tseytlin,
``Born-Infeld action, supersymmetry and string theory,''
arXiv:hep-th/9908105.
}

\lref\FelderSV{
G.~N.~Felder, L.~Kofman and A.~Starobinsky,
``Caustics in tachyon matter and other Born-Infeld scalars,''
JHEP {\bf 0209}, 026 (2002)
[arXiv:hep-th/0208019].
}

\lref\FelderXU{
G.~N.~Felder and L.~Kofman,
``Inhomogeneous fragmentation of the rolling tachyon,''
arXiv:hep-th/0403073.
}

\lref\SenXS{
A.~Sen,
``Open-closed duality at tree level,''
Phys.\ Rev.\ Lett.\  {\bf 91}, 181601 (2003)
[arXiv:hep-th/0306137].
}

\lref\SenIV{
A.~Sen,
``Open-closed duality: Lessons from matrix model,''
Mod.\ Phys.\ Lett.\ A {\bf 19}, 841 (2004)
[arXiv:hep-th/0308068].
}

\lref\SenNU{
A.~Sen,
``Rolling tachyon,''
JHEP {\bf 0204}, 048 (2002)
[arXiv:hep-th/0203211].
}

\lref\SenIN{
A.~Sen,
``Tachyon matter,''
JHEP {\bf 0207}, 065 (2002)
[arXiv:hep-th/0203265].
}

\lref\LarsenWC{
F.~Larsen, A.~Naqvi and S.~Terashima,
``Rolling tachyons and decaying branes,''
JHEP {\bf 0302}, 039 (2003)
[arXiv:hep-th/0212248].
}

\lref\OkudaYD{
T.~Okuda and S.~Sugimoto,
``Coupling of rolling tachyon to closed strings,''
Nucl.\ Phys.\ B {\bf 647}, 101 (2002)
[arXiv:hep-th/0208196].
}

\lref\LambertZR{
N.~Lambert, H.~Liu and J.~Maldacena,
``Closed strings from decaying D-branes,''
arXiv:hep-th/0303139.
}

\lref\GaiottoRM{
D.~Gaiotto, N.~Itzhaki and L.~Rastelli,
``Closed strings as imaginary D-branes,''
arXiv:hep-th/0304192.
}

\lref\SenMD{
A.~Sen,
``Supersymmetric world-volume action for non-BPS D-branes,''
JHEP {\bf 9910}, 008 (1999)
[arXiv:hep-th/9909062].
}

\lref\GarousiTR{
M.~R.~Garousi,
``Tachyon couplings on non-BPS D-branes and Dirac-Born-Infeld action,''
Nucl.\ Phys.\ B {\bf 584}, 284 (2000)
[arXiv:hep-th/0003122].
}

\lref\BergshoeffDQ{
E.~A.~Bergshoeff, M.~de Roo, T.~C.~de Wit, E.~Eyras and S.~Panda,
``T-duality and actions for non-BPS D-branes,''
JHEP {\bf 0005}, 009 (2000)
[arXiv:hep-th/0003221].
}

\lref\KlusonIY{
J.~Kluson,
``Proposal for non-BPS D-brane action,''
Phys.\ Rev.\ D {\bf 62}, 126003 (2000)
[arXiv:hep-th/0004106].
}

\lref\SenTM{
A.~Sen,
``Dirac-Born-Infeld action on the tachyon kink and vortex,''
Phys.\ Rev.\ D {\bf 68}, 066008 (2003)
[arXiv:hep-th/0303057].
}

\lref\SenMV{
A.~Sen,
``Remarks on tachyon driven cosmology,''
arXiv:hep-th/0312153.
}

\lref\ElitzurPQ{
S.~Elitzur, A.~Giveon, D.~Kutasov, E.~Rabinovici and G.~Sarkissian,
``D-branes in the background of NS fivebranes,''
JHEP {\bf 0008}, 046 (2000)
[arXiv:hep-th/0005052].
}

\lref\PelcKB{
O.~Pelc,
``On the quantization constraints for a D3 brane in the geometry of NS5
JHEP {\bf 0008}, 030 (2000)
[arXiv:hep-th/0007100].
}

\lref\RibaultSG{
S.~Ribault,
``D3-branes in NS5-branes backgrounds,''
hep-th/0301092,
JHEP {\bf 0302}, 044 (2003)
.
}

\lref\SeibergZK{
N.~Seiberg,
``New theories in six dimensions and matrix description of M-theory on  T**5
Phys.\ Lett.\ B {\bf 408}, 98 (1997)
[arXiv:hep-th/9705221].
}

\lref\DeWolfeQX{
O.~DeWolfe, S.~Kachru and H.~Verlinde,
``The giant inflaton,''
arXiv:hep-th/0403123.
}

\lref\AlishahihaEH{
M.~Alishahiha, E.~Silverstein and D.~Tong,
``DBI in the sky,''
arXiv:hep-th/0404084.
}

\lref\BuchelQG{
A.~Buchel and A.~Ghodsi,
``Braneworld inflation,''
arXiv:hep-th/0404151.
}

\lref\GutperleAI{
M.~Gutperle and A.~Strominger,
``Spacelike branes,''
JHEP {\bf 0204}, 018 (2002)
[arXiv:hep-th/0202210].
}

\lref\GibbonsMD{
G.~W.~Gibbons,
``Cosmological evolution of the rolling tachyon,''
Phys.\ Lett.\ B {\bf 537}, 1 (2002)
[arXiv:hep-th/0204008].
}

\lref\ShiuXP{
G.~Shiu, S.~H.~H.~Tye and I.~Wasserman,
``Rolling tachyon in brane world cosmology from superstring field theory,''
Phys.\ Rev.\ D {\bf 67}, 083517 (2003)
[arXiv:hep-th/0207119].
}

\lref\GiveonZM{
A.~Giveon, D.~Kutasov and O.~Pelc,
``Holography for non-critical superstrings,''
JHEP {\bf 9910}, 035 (1999)
[arXiv:hep-th/9907178].
}

\lref\AharonyUB{
O.~Aharony, M.~Berkooz, D.~Kutasov and N.~Seiberg,
``Linear dilatons, NS5-branes and holography,''
JHEP {\bf 9810}, 004 (1998)
[arXiv:hep-th/9808149].
}

\lref\GiveonPX{
A.~Giveon and D.~Kutasov,
``Little string theory in a double scaling limit,''
JHEP {\bf 9910}, 034 (1999)
[arXiv:hep-th/9909110].
}

\lref\BurgessQV{
C.~P.~Burgess, P.~Martineau, F.~Quevedo and R.~Rabadan,
``Branonium,''
JHEP {\bf 0306}, 037 (2003)
[arXiv:hep-th/0303170].
}

\lref\SilversteinHF{
E.~Silverstein and D.~Tong,
``Scalar speed limits and cosmology: Acceleration from D-cceleration,''
arXiv:hep-th/0310221.
}

\lref\KehagiasVR{
A.~Kehagias and E.~Kiritsis,
``Mirage cosmology,''
JHEP {\bf 9911}, 022 (1999)
[arXiv:hep-th/9910174].
}

\lref\BurgessTZ{
C.~P.~Burgess, F.~Quevedo, R.~Rabadan, G.~Tasinato and I.~Zavala,
``On bouncing brane worlds, S-branes and branonium cosmology,''
JCAP {\bf 0402}, 008 (2004)
[arXiv:hep-th/0310122].
}

\lref\LuninGW{
O.~Lunin, S.~D.~Mathur, I.~Y.~Park and A.~Saxena,
``Tachyon condensation and 'bounce' in the D1-D5 system,''
Nucl.\ Phys.\ B {\bf 679}, 299 (2004)
[arXiv:hep-th/0304007].
}

\lref\MaldacenaSS{
J.~M.~Maldacena, G.~W.~Moore and N.~Seiberg,
``D-brane charges in five-brane backgrounds,''
JHEP {\bf 0110}, 005 (2001)
[arXiv:hep-th/0108152].
}

\lref\GibbonsHF{
G.~W.~Gibbons, K.~Hori and P.~Yi,
``String fluid from unstable D-branes,''
Nucl.\ Phys.\ B {\bf 596}, 136 (2001)
[arXiv:hep-th/0009061].
}

\lref\GibbonsTV{
G.~Gibbons, K.~Hashimoto and P.~Yi,
``Tachyon condensates, Carrollian contraction of Lorentz group, and fundamental
strings,''
JHEP {\bf 0209}, 061 (2002)
[arXiv:hep-th/0209034].
}

\lref\YeeEC{
H.~U.~Yee and P.~Yi,
``Open / closed duality, unstable D-branes, and coarse-grained closed
Nucl.\ Phys.\ B {\bf 686}, 31 (2004)
[arXiv:hep-th/0402027].
}

\lref\SenMG{
A.~Sen,
``Non-BPS states and branes in string theory,''
arXiv:hep-th/9904207.
}

\lref\KutasovDJ{
D.~Kutasov,
``D-brane dynamics near NS5-branes,''
arXiv:hep-th/0405058.
}

\lref\NakayamaYX{
Y.~Nakayama, Y.~Sugawara and H.~Takayanagi,
``Boundary states for the rolling D-branes in NS5 background,''
JHEP {\bf 0407}, 020 (2004)
[arXiv:hep-th/0406173].
}

\lref\YavartanooWB{
H.~Yavartanoo,
``Cosmological solution from D-brane motion in NS5-branes background,''
arXiv:hep-th/0407079.
}

\lref\PanigrahiQR{
K.~L.~Panigrahi,
``D-brane dynamics in Dp-brane background,''
arXiv:hep-th/0407134.
}

\lref\GhodsiWN{
A.~Ghodsi and A.~E.~Mosaffa,
``D-brane Dynamics in RR Deformation of NS5-branes Background and Tachyon
arXiv:hep-th/0408015.
}

\lref\MinahanTG{
J.~A.~Minahan and B.~Zwiebach,
``Gauge fields and fermions in tachyon effective field theories,''
JHEP {\bf 0102}, 034 (2001)
[arXiv:hep-th/0011226].
}

\lref\AlishahihaDU{
M.~Alishahiha, H.~Ita and Y.~Oz,
``On superconnections and the tachyon effective action,''
Phys.\ Lett.\ B {\bf 503}, 181 (2001)
[arXiv:hep-th/0012222].
}

\lref\LosevHX{
A.~Losev, G.~W.~Moore and S.~L.~Shatashvili,
``M \& m's,''
Nucl.\ Phys.\ B {\bf 522}, 105 (1998)
[arXiv:hep-th/9707250].
}

\lref\OkuyamaWM{
K.~Okuyama,
``Wess-Zumino term in tachyon effective action,''
JHEP {\bf 0305}, 005 (2003)
[arXiv:hep-th/0304108].
}

\lref\HarveyWM{
J.~A.~Harvey, D.~Kutasov, E.~J.~Martinec and G.~Moore,
``Localized tachyons and RG flows,''
arXiv:hep-th/0111154.
}

\lref\sahakyan{D. Sahakyan, ``Comments on D-brane Dynamics Near NS5-branes,''
hep-th/0408070.}

\Title{
}
{\vbox{\centerline{A Geometric Interpretation }
\medskip
\centerline{of the Open String Tachyon}}}
\bigskip
\centerline{David Kutasov}
\bigskip
\centerline{{\it Enrico Fermi Inst. and Dept. of Physics,
University of Chicago}}
\centerline{\it 5640 S. Ellis Ave., Chicago, IL 60637-1433, USA}
\bigskip\bigskip\bigskip
\noindent
Unstable, non-BPS D-branes in weakly coupled ten dimensional string 
theory have many mysterious properties. Among other things, it is not 
clear what sets their tension,  what is their relation to the better 
understood BPS D-branes, and why the open string tachyon on them is 
described by an effective Lagrangian which suggests that the tachyon 
corresponds to an extra spatial dimension transverse to the branes. 
We point out that the dynamics of D-branes in the presence of 
Neveu-Schwarz fivebranes on a transverse $\IR^3\times S^1$ provides a 
useful toy model for studying these issues. From the point of view of 
a $5+1$ dimensional observer living on the fivebranes, BPS D-branes in 
ten dimensions give rise to two kinds of D-branes, which are BPS or 
non-BPS depending on whether they do or do not wrap the $S^1$. Their tensions
are related, since from a higher dimensional perspective, they are the
same objects. D-branes localized on the $S^1$ have a tachyon corresponding
to their position on the circle. This field is described by 
the same Lagrangian as that of a tachyon on a non-BPS D-brane in ten dimensions. 
Its geometrical interpretation is useful for clarifying the properties of 
non-BPS branes in six dimensions. If the lessons from the six dimensional 
system can be applied in ten dimensions, the existence of non-BPS D-branes 
seems to suggest the presence of at least one extra dimension in critical 
type II string theory.

\vfill

\Date{}

\newsec{Introduction}

In addition to the familiar BPS Dirichlet $p$-branes, which have even (odd)
$p$ in type IIA (IIB) string theory, the theory contains unstable, non-BPS
$Dp$-branes with  odd (even) $p$ in the IIA (IIB) case; see \eg\ \SenMG\ for 
a review. The non-BPS D-branes play an important role in string theory. In 
particular, wrapping them around various cycles often gives rise to stable, 
non-BPS states. They also appear as possible decay products in the process
of annihilation of BPS branes and anti-branes. Conversely, the BPS D-branes
can be thought of as solitons in the worldvolume theory of the non-BPS ones.

Despite much work on the properties of non-BPS D-branes, many basic
questions about them remain unanswered. Some of the questions that
motivated the present work are the following:
\item{(1)} Why do these branes exist? The existence of BPS D-branes
is natural since they are the lowest lying states for a given RR charge. 
The non-BPS branes do not carry RR charge: what fixes their tensions? 
\item{(2)} Is there a direct link between the non-BPS branes and the
BPS ones? What is the significance of the fact that whenever $p$ is
even for the BPS branes, it is odd for the non-BPS ones, and vice versa?
As mentioned above, the two types of branes are related by the
dynamical process of open string tachyon condensation (as well as
in other ways), which further suggests a close connection between them.
\item{(3)} The fact that non-BPS D-branes are unstable is reflected in the
presence of an open string tachyon $T$ in their spectrum. Many aspects
of the on-shell dynamics of this tachyon are captured by a spacetime effective
action of the Dirac-Born-Infeld (DBI) type\foot{Throughout this
paper we set to zero the gauge field on D-branes, and use the conventions
$\alpha'=1$ and $\eta_{\mu\nu}={\rm diag}(-1,+1,\cdots,+1)$.} \refs{\SenMD\GarousiTR
\BergshoeffDQ\KlusonIY\GibbonsHF\SenIN\SenAN\GibbonsTV\SenTM\LambertZR\KutasovER-\NiarchosRW},
\eqn\aaa{\SS_p=-\tau_p\int d^{p+1}x{1\over \cosh{T\over\sqrt2}}\sqrt{-\det G}~,}
where $\tau_p$ is the tension of the non-BPS D-brane, and
$G$ is the induced metric on the brane,
\eqn\bbb{G_{\mu\nu}=\eta_{\mu\nu}+\partial_\mu T\partial_\nu T
+\partial_\mu Y^I\partial_\nu Y^I ~.}
The scalar fields $Y^I$ ($I=p+1,\cdots,9$) living on the D-brane parametrize
its location in the transverse directions. The form of the induced metric \bbb\
seems to suggest that the tachyon direction in field space should be treated
as an extra dimension of space, like the $Y^I$. Can this be made more precise,
and if so, what is the meaning of the tachyon potential 
$V(T)=\tau_p/\cosh(T/\sqrt{2})$ in \aaa?
\item{(4)} Can one provide a geometric interpretation to the process
of the condensation of the open string tachyon $T$? In some other 
cases, an open string tachyon signals the fact that a  D-brane 
system can lower its energy by continuously changing its shape 
to a more stable configuration. An example of such a process is 
the reconnection of D-branes at an angle (see \eg\ fig. 13.4 in 
\PolchinskiRR). In the case of a non-BPS D-brane decaying into a 
lower dimensional BPS brane, 
there is no known geometric picture 
of the decay process, and it would be interesting to find one.

\noindent
In this note we will discuss a D-brane system which might help shed 
light on the questions raised above. The system involves
BPS D-branes propagating in the near-horizon geometry of Neveu-Schwarz
fivebranes. We will take the four dimensional space transverse to
the fivebranes to be $\IR^3\times S^1$. 

{}From the point of view of an observer living in the $5+1$ dimensional 
worldvolume of the fivebranes, BPS D-branes in the full ten dimensional
theory give rise to two kinds of finite tension D-branes in six dimensions. 
One consists of D-branes whose worldvolume lies entirely inside that of 
the fivebranes. Such D-branes are non-BPS -- while the fivebranes and the
D-branes separately preserve sixteen supercharges, a background that contains
both breaks all supersymmetry. The second kind consists of D-branes that 
wrap the circle transverse to the fivebranes. These D-branes preserve half 
of the supersymmetry of the fivebranes (eight supercharges) and are BPS. 

The situation here is reminiscent of that in ten dimensional string 
theory:
\item{(a)} In type IIA string theory, the $NS5$-branes preserve a chiral, 
$(2,0)$, supersymmetry in six dimensions (an analog of the chiral, IIB 
supersymmetry in ten dimensions). The BPS D-branes in the background of 
the fivebranes are $Dp$-branes with even $p$, wrapped around the transverse 
circle. From the six dimensional 
point of view, they have odd dimension. Similarly, the non-BPS D-branes are 
$Dp$-branes with even $p$ which are localized on the circle. Thus, we find 
the same pattern as in ten dimensional type IIB string theory: a chiral
supersymmetry, BPS D-branes of odd dimension and non-BPS D-branes of
even dimension. Both the BPS and the non-BPS D-branes
in six dimensions have the same higher dimensional origin -- they are BPS
D-branes in ten dimensions wrapped or unwrapped around the extra circle. 
This also explains why the dimensions of half of them (the BPS ones) are
odd, and of the other half even.
\item{(b)} In type IIB string theory, the story is similar. The fivebranes 
preserve a non-chiral, $(1,1)$ supersymmetry in six dimensions, an analog of
the supersymmetry of ten dimensional type IIA string theory, and the D-branes follow 
the same pattern as there: the (non-) BPS D-branes have  
(odd) even dimensions. 

\noindent
In section 2 we show that the analogy between the 
system of BPS D-branes propagating near $NS5$-branes, and the ten 
dimensional system of BPS and non-BPS branes is in fact much closer 
than one might have expected. The position of the branes
localized on the transverse circle (which as mentioned above are
non-BPS) behaves in a strikingly similar way to the tachyon on a
non-BPS D-brane in ten dimensions. At the same time, since it has 
a simple geometric interpretation, one can answer all the questions
raised above. In particular, as we show in section 3, the brane 
descent relations have a simple geometric interpretation.

The similarity to the original problem suggests that some of the 
lessons from this model can be applied to ten dimensional type II 
string theory as well. We briefly discuss this interesting possibility 
in section 4.

\newsec{D-branes near fivebranes on a transverse $\IR^3\times S^1$}

The dynamics of a BPS D-brane propagating in the vicinity of a stack of
$k$ $NS5$-branes on a transverse $\IR^4$ resembles that of an unstable 
D-brane in ten dimensions \KutasovDJ\ (see also 
\refs{\NakayamaYX\YavartanooWB\PanigrahiQR\GhodsiWN-\sahakyan} 
for further work along these
lines). Due to the gravitational attraction of the
D-brane to the fivebranes, the scalar field on the worldvolume of the
D-brane that corresponds to its distance from the fivebranes is described
by a Lagrangian very similar to \aaa, \bbb. The effective potential for
this field goes exponentially to zero at short distances (relative to
$\sqrt k$) and approaches a finite constant at large distances.

Thus, the  dynamics of the radial mode on a D-brane which is close to
a stack of fivebranes is very similar to that of  the tachyon on a
non-BPS D-brane. The behavior near the top of the potential is different --
while for the tachyon on a non-BPS brane the potential has a maximum at a
finite point in field space, for a D-brane propagating in the background
of fivebranes, the maximum of the potential is at infinity, and the form of
the potential near the maximum is different.

The main point of this note is that changing the space transverse to the fivebranes
from $\IR^4$ to $\IR^3\times S^1$, leads to a system whose dynamics is remarkably 
similar to that of BPS and non-BPS D-branes in ten dimensional type II string theory, 
and at the same time is well suited for studying the questions raised in \S1.

Consider a system of $k$ $NS5$-branes on a transverse $\IR^3\times S^1$,
which we will label by the coordinates $(\vec Z,Y)$, with $\vec Z\in \IR^3$
and $Y\sim Y+2\pi R$ ($R$ is the radius of the $S^1$). The fivebranes are
located at the point $\vec Z=Y=0$.  The background around them is\foot{There
is also a NS $B$-field, which will not play a role below.} (see \eg\ \PolchinskiRR)
\eqn\chs{\eqalign{
&ds^2=dx_\mu dx^\mu+H(\vec Z,Y)\left(d\vec Z^2+dY^2\right)~,\cr
&e^{2(\Phi-\Phi_0)}=H(\vec Z,Y)~.\cr
}}
Here $x^\mu\in \IR^{5,1}$ label the worldvolume of the fivebranes.
$\Phi_0$ is related to the string coupling far from the fivebranes,
$g_s=\exp\Phi_0$. The harmonic function $H$ in \chs\ has the form
\eqn\harfun{H=1+k\sum_{n=-\infty}^\infty{1\over (Y-2\pi Rn)^2+\vec Z^2}~.}
We will be interested in studying the system in the near-horizon limit,
which can be defined by rescaling all distances by a factor of $g_s$,
\eqn\rescal{Y=g_sy;\;\;\vec Z=g_s \vec z;\;\;R=g_s r~,}
and sending $g_s\to 0$ while keeping the rescaled distances $(y,\vec z,r)$
fixed. This leads to the background
\eqn\chsnew{\eqalign{
&ds^2=dx_\mu dx^\mu+h(\vec z,y)\left(d\vec z^2+dy^2\right)~,\cr
&e^{2\Phi}=h(\vec z,y)~,\cr
}}
with
\eqn\harfunnew{h=k\sum_{n=-\infty}^\infty{1\over (y-2\pi rn)^2+\vec z^2}
={k\over 2r|\vec z|}{\sinh(|\vec z|/r)\over\cosh(|\vec z|/r)-\cos(y/r)}~.}
When the space transverse to the fivebranes is $\IR^4$, an analogous limit 
is usually taken to study the dynamics on the fivebranes given by Little 
String Theory \AharonyUB. It has been argued in \LosevHX\ that when the 
transverse space includes an $S^1$, one can also decouple the fivebranes 
from the bulk. This interesting issue will not be discussed here.

Since the radius of the circle, $R$ \rescal, goes to zero in the limit
under consideration, it is natural to ask whether performing T-duality 
along the circle leads to a better description of the physics. In fact, 
the limit taken here is very similar to the one relevant for the T-duality
between string theory on $\IR^{5,1}\times \IC^2/Z_n$ and its description
in terms of fivebranes on a transverse $\IR^3\times S^1$. As reviewed 
in section 4.2 of \HarveyWM, the perturbative orbifold is T-dual to a
background which differs from the one considered here only in the fact 
that the fivebranes are spread at equal distances on the $y$ circle, rather
than placed at the same point. Thus, our background can be alternatively 
described by starting with the perturbative orbifold and turning off the 
$B$-fields through the $n-1$ minimal cycles.

This T-dual orbifold description {\it is} more useful when the fivebranes are
spread around the circle, but it is {\it not} useful in our background.
The reason is that while $Y$ naively lives on a very small circle, near the
location of the fivebranes ($Y=0$) an infinite throat develops, and the metric
\chs\ becomes $ds^2=kdY^2/Y^2+\cdots$. In particular, rather than being vanishingly
small, the distance to $Y=0$ is infinite.

We now place a D-brane whose worldvolume is embedded entirely in $\IR^{5,1}$,
in the geometry \chsnew, \harfunnew. The varying dilaton provides a potential
for the D-brane, which attracts it to the fivebranes. An unstable equilibrium 
can be achieved by placing the D-brane diametrically opposite the fivebranes 
on the $S^1$, at $\vec z=0$, $y=\pi r$, where the total force on it vanishes.
Expanding around this point, it is clear that the scalar field $y(x^\mu)$
on the D-brane is tachyonic, since the energy of the D-brane decreases
when it moves towards the fivebranes located at $y=0$. The three fields
$\vec z$ are massive -- the string coupling decreases, and thus the energy
of the D-brane increases, as it moves away from the origin in $\IR^3$. 
The mass of $\vec z$ can be calculated from the DBI action; one finds
\eqn\mmzz{M_z=1/\sqrt{3k}~.}

\noindent
We see that a BPS D-brane placed opposite a stack of $NS5$-branes 
on a transverse circle has similar qualitative properties to a
non-BPS D-brane in ten dimensional flat spacetime. The $NS5$-brane
background, which is the analog of ten dimensional flat spacetime,
preserves six dimensional Poincare symmetry and 
sixteen supercharges. 
Adding the D-brane breaks all supersymmetry. The spectrum of open 
strings on the D-brane contains one tachyonic scalar field, which 
in our case corresponds to $y(x^\mu)$ and for the non-BPS brane in 
ten dimensions is the open string tachyon $T$ \aaa.
Condensation of this tachyon leads to the annihilation of the unstable
D-brane. For the D-NS system, the D-brane falls onto the fivebranes and is 
absorbed by them. This process is discussed in more detail in \KutasovDJ.

It is important to emphasize that the analog of the ten dimensional
flat spacetime of type II string theory in the fivebrane case is the
six dimensional worldvolume of the fivebranes. In other words, we are
taking the point of view of an observer living on the fivebranes.\foot{As
is familiar from discussions of Little String Theory.} From
this point of view, $y(x^\mu)$, $\vec z(x^\mu)$ are non-geometrical fields
on the D-brane, and not extra directions of space, like the tachyon and
massive modes living on a non-BPS D-brane in ten dimensional flat spacetime.

The D-NS system is useful because in it one can take a ``broader,''
ten dimensional, perspective, in which these fields are geometrical and
their dynamics is easier to visualize. This is especially instructive since, as
we show next, the two systems are much more closely related than one might
have expected.

Our main interest here will be in the dynamics of the tachyon $y(x^\mu)$
on the unstable D-brane described above. The DBI action for this field
is given by
\eqn\dbiradial{\SS_p=-T_p\int d^{p+1}x{1\over\sqrt h}
\sqrt{1+h(y)\partial_\mu y\partial^\mu y}~.}
Here $T_p$ is defined such that the tension of a BPS D-brane in flat spacetime
is $T_p/g_s$. The harmonic function $h(y)$ is obtained by placing the massive
fields $\vec z$ at the minimum of their potential, $\vec z=0$, so that \harfunnew\
reduces to
\eqn\hyy{h(y)={k\over4r^2\sin^2{y\over 2r}}~.}
In particular, setting $y=\pi r$ we see that the tension of the unstable D-brane
is
\eqn\tenunst{\tau_p^{\rm (unstable)}={2T_pR\over\sqrt{k}g_s}~.}
Note that the tension \tenunst\ is proportional to the radius of the circle
$R$, despite the fact that the D-brane is localized on it.

As in \KutasovDJ,  we would like to bring the action \dbiradial\ to the form \aaa.
To this end, we define a ``tachyon'' field $T$ via the relation
\eqn\tachdef{{dT\over dy}=\sqrt{h(y)}={\sqrt{k}\over2r\sin{y\over2r}}~,}
whose solution is
\eqn\solt{\sinh{T\over\sqrt{k}}=-\cot{y\over2r}~.}
This field redefinition brings the argument of the square root in \dbiradial\
to the form \aaa. The ``tachyon potential'' in \dbiradial\ is given by the
prefactor, $T_p/\sqrt h$. Rewriting it in terms of the field $T$ \solt, we find
\eqn\tachpoten{V(T)={\tau_p^{\rm (unstable)}\over\cosh{T\over\sqrt{k}}}~.}
Thus, the full DBI Lagrangian for the unstable D-brane obtained by placing
a BPS D-brane on the opposite side of the circle from $k$ $NS5$-branes
is given by
\eqn\unstlag{\SS_p=-\tau_p^{\rm (unstable)}
\int d^{p+1}x{1\over \cosh{T\over\sqrt k}}\sqrt{-\det G}~,}
where the induced metric is given again by \bbb\ and the ``tachyon'' $T$
parametrizes the position of the D-brane on the $y$ circle via \solt. The
mass of the tachyon is obtained by expanding \unstlag\ around $T=0$:
\eqn\mtk{M_T^2=-{1\over k}~.}

\noindent
Note that the Lagrangian \unstlag\ has almost the same form as that of an unstable 
D-brane in type II string theory in ten dimensions \aaa. For $k=2$ (two 
fivebranes), the two Lagrangians \aaa, \unstlag\ are identical.\foot{In 
\KutasovDJ\ it was noted that for fivebranes on a transverse $\IR^4$, for 
$k=2$ one finds the same large $T$ behavior of the potential in the two 
problems. Now we see that on $\IR^3\times S^1$ the two Lagrangians coincide 
for all $T$. In particular, the mass of the open string tachyon on a non-BPS 
D-brane in ten dimensions is the same as the mass of the tachyonic mode 
corresponding to translations along the $y$ circle \mtk\ for a BPS D-brane 
moving in the background of two fivebranes.} 

In \KutasovER\ it was emphasized that the tachyon DBI Lagrangian \aaa, \bbb\ 
is closely related to the usual DBI one. The construction of \KutasovDJ\
and this section is a manifestation of this relation. We found the tachyon
DBI Lagrangian by starting with the standard DBI one in a non-trivial
gravitational potential.

\newsec{The geometric interpretation of kinks on unstable D-branes}

In the previous section we saw that a BPS D-brane placed on a circle in
the vicinity of $k$ $NS5$-branes has a tachyon corresponding to its position 
on the circle, and that this tachyon is described by essentially the same 
Lagrangian as that of a non-BPS D-brane in ten dimensional string theory. 
In the latter context there are known solutions of the equations of motion 
that describe lower dimensional D-branes as kinks on the worldvolume of 
the non-BPS D-brane. In this section we will describe the geometric 
interpretation of these solutions in our problem.  

In ten dimensions, one can construct a BPS $D(p-1)$-brane as a kink on the worldvolume 
of a non-BPS $Dp$-brane  \refs{\SenMG,\KlusonIY,\MinahanTG,\AlishahihaDU,\SenTM}.
The kink is a static solution of the equations of motion of the tachyon DBI action 
\aaa, for which $T$ depends on a single spatial coordinate, say $x^p$, and satisfies
the boundary conditions $T\to\pm\infty$ as $x^p\to\pm\infty$. The kink is infinitely thin:
$T=-\infty$ for all $x^p<x_{\rm kink}$ and $T=+\infty$ for all $x^p>x_{\rm kink}$.
Its tension is given in terms of the tachyon potential by
\eqn\tpminusone{\tau_{p-1}=\int_{-\infty}^\infty V(T)dT~.}
One of the successes of the tachyon DBI action \aaa\ is that the result 
\tpminusone\ agrees exactly with the full string theory brane descent relation. 

It is natural to apply this construction to the unstable D-brane discussed in the
previous section. Plugging in the potential \tachpoten\ into \tpminusone, and using
\tenunst, one finds that in our case, the tension of the kink is
\eqn\taupmo{\tau_{p-1}=\tau_p^{\rm (unstable)}\pi\sqrt{k}={2\pi RT_p\over g_s}~.}
The geometrical meaning of this kink is clear. Since the tachyon $T$ is directly 
related in our problem to position on the $y$ circle, \solt, the kink solution 
corresponds to a D-brane which sits on top of the fivebranes,\foot{Recall that 
a D-brane sitting on top of the fivebranes has energy of order one in string units, 
\ie\ in classical open string theory it is like no D-brane at all. See \KutasovDJ\ 
for further discussion.} at $y=0$, for all $x^p<x_{\rm kink}$, then at $x^p=x_{\rm kink}$ 
it goes around the $y$ circle and back to the fivebranes at $y=2\pi r$ (which is the 
same as $y=0$), where it stays for all $x^p>x_{\rm kink}$. Thus, it describes a BPS
D-brane wrapped around the $y$ circle. The tension of the kink \taupmo\ is indeed 
precisely equal to that of  such a brane. 

As in the case of the unstable D-brane in ten dimensions, the D-brane 
described by the kink is supersymmetric. In our case this is the statement
that while a system consisting of $NS5$-branes stretched in the directions
$(x^1,x^2,x^3,x^4,x^5)$ and a $Dp$-brane stretched in $(x^1,\cdots,x^p)$,
breaks supersymmetry completely, a system that contains the above fivebranes
and a $Dp$-brane stretched in $(x^1,\cdots, x^{p-1},y)$, where $y$ is one of
the directions transverse to the fivebranes, preserves eight of the sixteen
supercharges which are preserved by the fivebranes alone. 

Note that both of the objects seen by a six dimensional observer living on 
the fivebranes as a non-BPS $Dp$-brane and a BPS $D(p-1)$-brane,
are the same type of object in the underlying ten dimensional theory.
Both are BPS $Dp$-branes oriented in different ways on $\IR^{5,1}\times S^1$.
This explains why the BPS D-brane can appear as a kink on the worldvolume of
a non-BPS one.

The fact that the D-brane described by the tachyon kink is supersymmetric
means that in addition to \unstlag, the tachyon Lagrangian must include a 
Wess-Zumino term which couples the tachyon to the appropriate RR gauge field. 
Of course, since the original ``non-BPS brane'' corresponds from the ten
dimensional perspective to a BPS $Dp$-brane at a point on the $y$ circle, 
it couples to the $(p+1)$-form RR gauge field $A^{(p+1)}$ in six dimensions, 
via a coupling of the form 
\eqn\wzone{\int_{V_{p+1}}A^{(p+1)}~.}
However, the charge corresponding to $A^{(p+1)}$ is not conserved in the
fivebrane background (see \KutasovDJ\ for some further comments on this),
and we will not discuss it in detail here.

The RR gauge field that is of more interest is the $p$-form $C^{(p)}$
in $5+1$ dimensions, obtained by taking the $(p+1)$-form $A^{(p+1)}$ with 
one of the indices being $y$, \ie\ $A^{(p+1)}=C^{(p)}\wedge dy$. This is 
the RR field that couples to a BPS $Dp$-brane which looks like a 
$(p-1)$-brane in $5+1$ dimensions and is further wrapped around the $y$ 
circle. Plugging this into \wzone, we find that the Wess-Zumino term in 
question goes like
\eqn\wztwo{\int_{V_{p+1}}C^{(p)}\wedge dy~.}
Using the relations \tachdef-\tachpoten, one finds that it is proportional to 
\eqn\wzthree{\int_{V_{p+1}}C^{(p)}\wedge dT V(T)~,}  
where $V(T)$ is the ``tachyon potential'' \tachpoten. We will not 
compute the overall normalization of \wzthree\ here (although it
would be useful to do so), since it is
completely determined by the tension \taupmo\ and the requirement 
that the kink on the non-BPS brane should correspond to a BPS brane wrapped 
around the $y$ circle. What is interesting is that the $T$ dependence 
of the Wess-Zumino term \wzthree\ is precisely the same as that for a 
non-BPS D-brane in ten dimensional string theory (the latter was 
computed in \OkuyamaWM). This is one more indication of the close 
relation between the dynamics of D-branes near $NS5$-branes, and the
dynamics of non-BPS branes in ten dimensional string theory.

In the D-NS system, the time-dependent decay of a non-BPS $Dp$-brane to
a BPS $D(p-1)$-brane has a simple geometric interpretation. Suppose we start
with a D-brane which is stretched in a direction along the fivebranes, $x^p$,
and is unstably balanced at $y=\pi r$. If we displace it from $y=\pi r$ 
uniformly in $x^p$, the D-brane will ``roll'' towards the fivebranes and
annihilate (as in \KutasovDJ). If, on the other hand, we start with a 
configuration where $y(x^p)<\pi r$ for $x_p<x_{\rm kink}$ and $y(x^p)>\pi r$ 
for $x_p>x_{\rm kink}$, it is plausible that the part of the D-brane with 
$x_p<x_{\rm kink}$ will approach $y=0$ at late times\foot{In general, one 
would expect the final value of $x_{\rm kink}$ to be different from its
initial value.}, while the part with $x_p>x_{\rm kink}$ will approach 
$y=2\pi r$. The parts of the D-brane at $y=0,2\pi r$ will dissolve in the 
fivebranes, and only the part at $x^p=x_{\rm kink}$, which wraps once around 
the $y$ circle, will remain. From the six dimensional point of view, this time 
evolution describes the decay of a non-BPS $Dp$-brane to a BPS $D(p-1)$-brane.

\newsec{Discussion}

The main result of this note is the observation that the dynamics of D-branes,
which are BPS in flat ten dimensional spacetime, propagating
in the background of $NS5$-branes  on a transverse $\IR^3\times S^1$,
is remarkably similar to that of BPS and non-BPS D-branes in ten dimensions. 
At the same time, the questions regarding non-BPS branes listed in \S1 can be
answered in this system. Let us summarize the answers. 

The answer to questions (1) and (2) is that what looks to a six dimensional
observer like BPS and non-BPS branes, are in fact the same type of object --
BPS D-branes in ten dimensions wrapped or unwrapped around the extra $S^1$.
This explains why the non-BPS branes exist, why their dimensions differ by
one from the BPS ones and their tensions take the values they do, and why 
they do not carry (conserved) RR charges. 

The answer to question (3) is that the form of the induced metric \bbb\ indeed
indicates in this case that the tachyon parametrizes an extra dimension of space.
It is related to the coordinate on the circle, $y$, via \solt. The tachyon potential
\tachpoten\ is due to a gravitational source (the $NS5$-branes)
localized on the $S^1$. The Wess-Zumino term on the non-BPS brane,
\wzthree, has a simple interpretation, as the pullback of \wztwo\ to the
worldvolume of the unstable D-brane.

Finally, the answer to question (4) is positive. The homogeneous 
``rolling tachyon'' solution describes a D-brane falling towards
the fivebranes, as in \KutasovDJ. Tachyon condensation to lower
dimensional branes corresponds to inhomogeneous solutions, where
different parts of the D-brane fall in different directions.

A natural question is whether all this teaches us anything about the
physics of BPS and non-BPS D-branes in ten dimensional flat spacetime.
Here, the situation is much less clear. For example, we found that the
Lagrangian describing D-branes localized on the $S^1$ in the presence of
$k$ coincident fivebranes is very similar, and for two fivebranes identical, to that of
non-BPS D-branes in type II string theory in flat spacetime. This includes
both the DBI terms \aaa, \unstlag, and the Wess-Zumino term \wzthree. 

We do not have a satisfactory explanation of this fact. The two Lagrangians
were arrived at in different ways (although the basic philosophy leading
to them is similar \KutasovER) and, more importantly, they describe different
systems. It is not clear whether there is a deeper connection between these
problems.\foot{It is perhaps worth pointing out another,
seemingly independent, relation between the system
of $k=2$ fivebranes and ten dimensional string theory. 
The Hagedorn temperature of the fivebrane theory (or 
Little String Theory) is in this case the 
same as that of perturbative ten dimensional 
string theory.}

Whether or not there is a direct relation between the dynamics of D-branes
near fivebranes and non-BPS branes in flat spacetime, it is natural to ask
whether some or all of the lessons from the six dimensional problem can be
used in the ten dimensional one. 

In ten dimensions, one can run the construction of this note in reverse.
Start from the DBI action \aaa, and Wess-Zumino term \wzthree\ for non-BPS
D-branes in ten dimensions, define $y$ using \solt\ (with $k=2$). This
brings the DBI action to the form \dbiradial, and the Wess-Zumino term 
to the form \wztwo. It seems tempting to postulate that $y$ should again
be interpreted as an extra spatial dimension (in addition to the nine usual
ones), and $h(y)$ \hyy\ is a gravitational potential due to a source at
$y=0$. The BPS D-branes would presumably again be wrapped around the $y$
direction, which would thus have to be compact, as in our six dimensional
example. 

To make this picture concrete and useful requires a better
understanding of  the extra dimension parametrized by $y$, 
and the source of the gravitational potential corresponding to $h(y)$.

\bigskip
\noindent{\bf Acknowledgements:}
I am grateful to A. Parnachev, D. Sahakyan and S. Shenker for
useful discussions. I also thank the Aspen Center for Physics 
for hospitality. This work was supported in part by DOE grant 
\#DE-FG02-90ER40560.

\listrefs
\end